\documentclass[conference]{IEEEtran}
\IEEEoverridecommandlockouts
\usepackage{cite}
\usepackage{amsmath,amssymb,amsfonts}
\usepackage{algorithmic}
\usepackage{graphicx}
\usepackage{textcomp}
\usepackage{xcolor}
\usepackage{booktabs}
\def\BibTeX{{\rm B\kern-.05em{\sc i\kern-.025em b}\kern-.08em
    T\kern-.1667em\lower.7ex\hbox{E}\kern-.125emX}}
\begin{document}

\title{VC-1K: A Diverse Video Test Set for Real-Time Communication}

\author{\IEEEauthorblockN{Yu-Chen Sun, Jie Dong, Sergey Sablin, Hassene Tmar, Sriram Srinivasan}
\IEEEauthorblockA{\textit{Meta} \\
\{ycsun, jiedong, sergeysablin, htmar, sriramsri\}@meta.com}
}

\maketitle

\begin{abstract}
We introduce Video Conferencing 1000 (VC-1K), an open-source test set comprising approximately 1,000 video sequences specifically designed for evaluating video processing and encoding pipelines in real-time communication (RTC) scenarios. Unlike existing test sets that primarily rely on studio-produced or screen content, VC-1K consists of user-generated video call content captured on consumer mobile devices under a wide variety of conditions. The dataset predominantly features talking-head sequences with distinct faces, but also includes a small subset of outward-facing, rear-camera videos without faces to reflect additional common RTC use cases. It spans a wide range of real-world variables including device type, lighting conditions, motion levels, indoor/outdoor environments, and demographic diversity. All video sequences are accompanied by synchronous audio to facilitate multimodal research. We detail the dataset's design methodology and statistical distribution, and demonstrate its utility through a comparative encoder evaluation of OpenH264, SVT-AV1, and libaom under constant bitrate (CBR) encoding at resolutions typical of RTC deployments. Furthermore, two super-resolution approaches, RealESRGAN and GFPGAN, are evaluated on VC-1K, highlighting the dataset as a useful benchmark for pre- and post-processing research. VC-1K is intended to serve as a publicly available dataset for the video RTC research community.
\end{abstract}


\section{Introduction}
Real-time video communication (RTC) has become ubiquitous with applications ranging from video calling and teleconferencing to interactive live streaming. RTC video has distinct characteristics, predominantly composed of user-generated talking-head sequences, captured on mobile devices across indoor and outdoor environments, and subject to both facial and camera motion. Optimizing the underlying components of these systems such as video codecs and pre/post-processing algorithms has a measurable impact on user experience.

Video codecs, which balance compression efficiency against strict latency and compute constraints, are critical to delivering a good quality of experience in RTC applications. However, evaluating these technologies effectively remains a challenge. Despite mature standards and next-generation codecs such as AV1, no public test set accurately represents RTC video, especially user-generated face videos. Existing benchmarks such as the AOM \cite{b1} or JVET \cite{b2} Common Test Conditions (CTC) or professional broadcast content target high-bitrate offline encoding or are limited in quantity, missing the key attributes of RTC video: faces, mobile front-facing capture, variable lighting, user motion, and constrained resolutions.

To address this gap, we present VC-1K, a curated dataset of $\sim$1,000 sequences representing RTC content, collected from diverse participants across multiple countries using their own mobile devices in natural environments. Each video is annotated with rich metadata: device, resolution, frame rate, lighting, motion level, environment, camera orientation, and participant demographics.

We demonstrate its utility through a comparative evaluation of three widely-used software encoders, OpenH264 (H.264/AVC), SVT-AV1, and libaom (AV1), under constant bitrate (CBR) control at 360p, mirroring the offline benchmarking methodology used in production video-calling deployments. We measure compression efficiency via BD-Rate \cite{b3} and complexity via CPU user/system time.

Pre/post-processing also meaningfully improves RTC visual quality, but ML-based methods such as super-resolution depend on data representative of real RTC scenarios. VC-1K addresses this requirement. By benchmarking two prominent super-resolution models \cite{b4,b5} on our dataset, we expose critical quality gaps unique to mobile video calls. This demonstrates the utility and challenging nature of VC-1K, highlighting it as a rigorous benchmark that reveals concrete opportunities for future algorithm design.

VC-1K's synchronized audio and face video under diverse real-world conditions enables evaluation of audio-visual models in realistic RTC scenarios that existing datasets \cite{b6,b7,b8} do not capture. VC-1K is best suited as an evaluation benchmark complementing larger training corpora. Three directions are promising. Visual voice activity detection \cite{b9} can be benchmarked against VC-1K's low-light and high-motion conditions across diverse device cameras and frame rates (30/24 fps). Audio-visual speech enhancement \cite{b7,b10}, where lip features guide noise suppression, can be evaluated under realistic visual degradations such as H.264 compression artifacts, low lighting, and motion blur that critically affect visual feature extraction in production but are absent from studio corpora. VC-1K's audio can be further augmented with noise from the DNS Challenge \cite{b11}. Lip synchronization metrics such as SyncNet \cite{b12} can be tested under real mobile-induced latencies, variable frame rates, and encoding artifacts, conditions absent from broadcast-derived benchmarks \cite{b13}.

The contributions of this paper are:
\begin{enumerate}
\item A publicly available, diverse video dataset tailored for the evaluation of end-to-end RTC pipelines, including encoding and pre-/post-processing algorithms;
\item A detailed characterization of the dataset's statistical properties, real-world environmental variables, and demographic diversity;
\item Comprehensive baseline benchmarks covering both open-source encoder evaluations and machine learning-based super-resolution models under realistic RTC operating conditions using open-source evaluation scripts \cite{b14} for full reproducibility.
\end{enumerate}

\section{Related Work}
\textbf{\textit{Standard Test Conditions.}} The AOM Common Test Conditions (CTC) v9.0 \cite{b1} and JVET CTC \cite{b2} define standardized sequences for AV2 and VVC/H.266 development, including low-delay modes relevant to conferencing. However, both rely on professionally captured, high-resolution content, with little representation of consumer mobile video-calling captures.

\textbf{\textit{User-Generated Content Datasets.}} YouTube-UGC \cite{b15} offers 1,500 UGC clips for compression research but without controlled metadata or RTC-specific characteristics. LIVE-VQC \cite{b16} provides 585 videos with authentic in-capture distortions from 101 devices, but targets no-reference quality assessment rather than codec benchmarking and lacks RTC-specific talking-head content.

\textbf{\textit{Facial Video Content Datasets.}} VoxCeleb \cite{b17}, VoxCeleb2 \cite{b8}, and YouTube Faces \cite{b18} provide large-scale talking-head collections for speaker/face recognition, but are sourced mainly from professional cameras or high-quality web uploads, not the heterogeneous devices and authentic capture conditions of production RTC. VCD \cite{b19} is closer in scope, offering a video-conferencing dataset explicitly intended for compression research. Its captures, however, are predominantly desktop and laptop webcam recordings collected under largely favorable lighting conditions, so mobile-device content and the more challenging capture conditions typical of mobile RTC remain comparatively underrepresented.

\textbf{\textit{Other Benchmarking Resources.}} UVG \cite{b20}, Netflix Chimera, and Xiph.org / ITU-T P-series sequences offer high-quality streaming/broadcast content far removed from mobile video calling.

In contrast, VC-1K provides a structured, annotated, publicly accessible dataset built explicitly for the video-communication use case, with controlled diversity across device, environment, lighting, motion, and demographics.

\section{Dataset Description}
\subsection{Collection Methodology}
The VC-1K dataset was collected through a structured data collection program engaging participants across five countries: the United States, Brazil, Mexico, India, and the Philippines. Participants were recruited to represent diverse demographics and were instructed to record short video sequences ($\sim$15 seconds) using their personal mobile devices in their natural environments (homes and public spaces).

The collection protocol specified controlled variation across device, camera, orientation, lighting, motion, and environment, with all sequences captured at 720p (1280$\times$720 landscape / 720$\times$1280 portrait).

All videos were captured using the device's native camera encoder (H.264/AVC) by the default high quality settings. We checked all the video and confirmed visually lossless quality, ensuring high source fidelity suitable for codec evaluation. The videos in the dataset were also recorded with voice. Participants provided informed consent and signed research participation agreements prior to data collection.

\subsection{Dataset Characteristics}
The final dataset comprises 1,007 video sequences (930 with consistent frame rates and 77 exhibiting frame rate jitter from device variability). Table~\ref{tab:attr} summarizes the distribution of key attributes.

\begin{table}[t]
\caption{VC-1K Dataset Attribute Distribution. Video Breakdown: 1007 Total (899 With Faces; 108 Outward-Facing, Rear-Camera, No-Face Videos)}
\label{tab:attr}
\centering
%\footnotesize
\begin{tabular}{@{}llrr@{}}
\toprule
\textbf{Attribute} & \textbf{Category} & \textbf{Count} & \textbf{Percentage} \\
\midrule
\textbf{Device}       & Android      & 798       & 79.2\% \\
                      & iOS          & 209       & 20.8\% \\ \addlinespace
\textbf{Orientation}  & Portrait     & 907       & 90.1\% \\
                      & Landscape    & 100       & 9.9\%  \\ \addlinespace
\textbf{Camera}       & Front-facing & 898       & 89.2\% \\
                      & Rear-facing  & 109       & 10.8\% \\ \addlinespace
\textbf{Environment}  & Indoor       & 760       & 75.5\% \\
                      & Outdoor      & 247       & 24.5\% \\ \addlinespace
\textbf{Motion}       & Slow         & 569       & 56.5\% \\
                      & Medium       & 303       & 30.1\% \\
                      & Fast         & 135       & 13.4\% \\ \addlinespace
\textbf{Lighting}     & Good         & 577       & 57.3\% \\
                      & Medium       & 347       & 34.5\% \\
                      & Low          & 59        & 5.9\%  \\
                      & Dark         & 24        & 2.4\%  \\ \addlinespace
\textbf{Frame Rate}   & 30 fps       & $\sim$96\% & ---    \\
                      & 24/25 fps    & $\sim$4\%  & ---    \\
\bottomrule
\end{tabular}
\end{table}

\subsection{Demographic Diversity}
A key design goal of VC-1K is demographic representativeness. Unlike many existing datasets that skew toward specific populations, VC-1K was collected with explicit diversity targets across age, gender, ethnicity, and geography.

\begin{table}[htb]
\caption{Demographic Distribution. This Table Includes 899 Videos With Distinct Faces and Excludes 108 No-Talking-Head Videos.}
\label{tab:demo}
\centering
%\footnotesize
\begin{tabular}{@{}llrr@{}}
\toprule
\textbf{Attribute} & \textbf{Category} & \textbf{Count} & \textbf{Percentage} \\
\midrule
\textbf{Gender}    & Female          & 452 & 50.3\% \\
                   & Male            & 445 & 49.5\% \\
                   & Non-binary      & 2   & 0.2\%  \\ \addlinespace
\textbf{Age}       & 18--24          & 231 & 25.7\% \\
                   & 25--50          & 445 & 49.5\% \\
                   & 51+             & 223 & 24.8\% \\ \addlinespace
\textbf{Ethnicity} & Black           & 227 & 25.3\% \\
                   & White           & 224 & 24.9\% \\
                   & Latino/Hispanic & 221 & 24.6\% \\
                   & South Asian     & 130 & 14.5\% \\
                   & Southeast Asian & 94  & 10.5\% \\
                   & East Asian      & 3   & 0.3\%  \\ \addlinespace
\textbf{Country}   & United States   & 278 & 30.9\% \\
                   & Brazil          & 259 & 28.8\% \\
                   & Mexico          & 138 & 15.4\% \\
                   & India           & 131 & 14.6\% \\
                   & Philippines     & 93  & 10.3\% \\
\bottomrule
\end{tabular}
\end{table}

This diversity ensures that technologies are evaluated across a representative range of skin tones, facial features, and environmental contexts---factors that can impact encoder behavior in video communication scenarios, particularly for face detection-based region-of-interest video coding.

\subsection{Content Characteristics}
The majority of sequences (89\%) feature a front-facing talking head, representative of the dominant video calling modality. The remaining sequences feature rear-camera footage capturing environments and objects, representing screen-sharing or environmental-show scenarios common in video calls.

Motion levels are intentionally skewed toward slow motion (57\%), reflecting typical video calling behavior where participants are relatively stationary, with medium (30\%) and fast (13\%) motion sequences providing challenging test cases for encoder motion estimation.

\section{Encoder Evaluation}

We evaluate three open-source RTC encoders, OpenH264 (H.264/AVC) \cite{b21}, SVT-AV1 \cite{b22}, and libaom (reference AV1) \cite{b23}, configured for real-time communication with constant bitrate (CBR) rate control, single-pass encoding without lookahead, at resolutions of 360p across codec-specific speed presets (OpenH264 0--1, SVT-AV1 9--11, libaom 9--11). Compression efficiency is measured using Y-PSNR averaged per sequence, with Bj\o ntegaard Delta Rate (BD-Rate) computed against OpenH264 speed 0 as the anchor. Computational complexity is assessed via user time (CPU encoding cycles) and system time (kernel overhead) in single-threaded mode. All experiments leverage the SVT-AV1 repository's open-source automation scripts \cite{b14} for encoding, quality assessment, and BD-Rate calculation, ensuring full reproducibility and extensibility.

Benchmarking was performed on 360p, the typical resolution used in video conferencing, downscaled images using ffmpeg lanczos filter, host system is Mac M1 Ultra 2022.

\subsection{Results}
\begin{figure}[htbp]
\centerline{\includegraphics[width=\columnwidth]{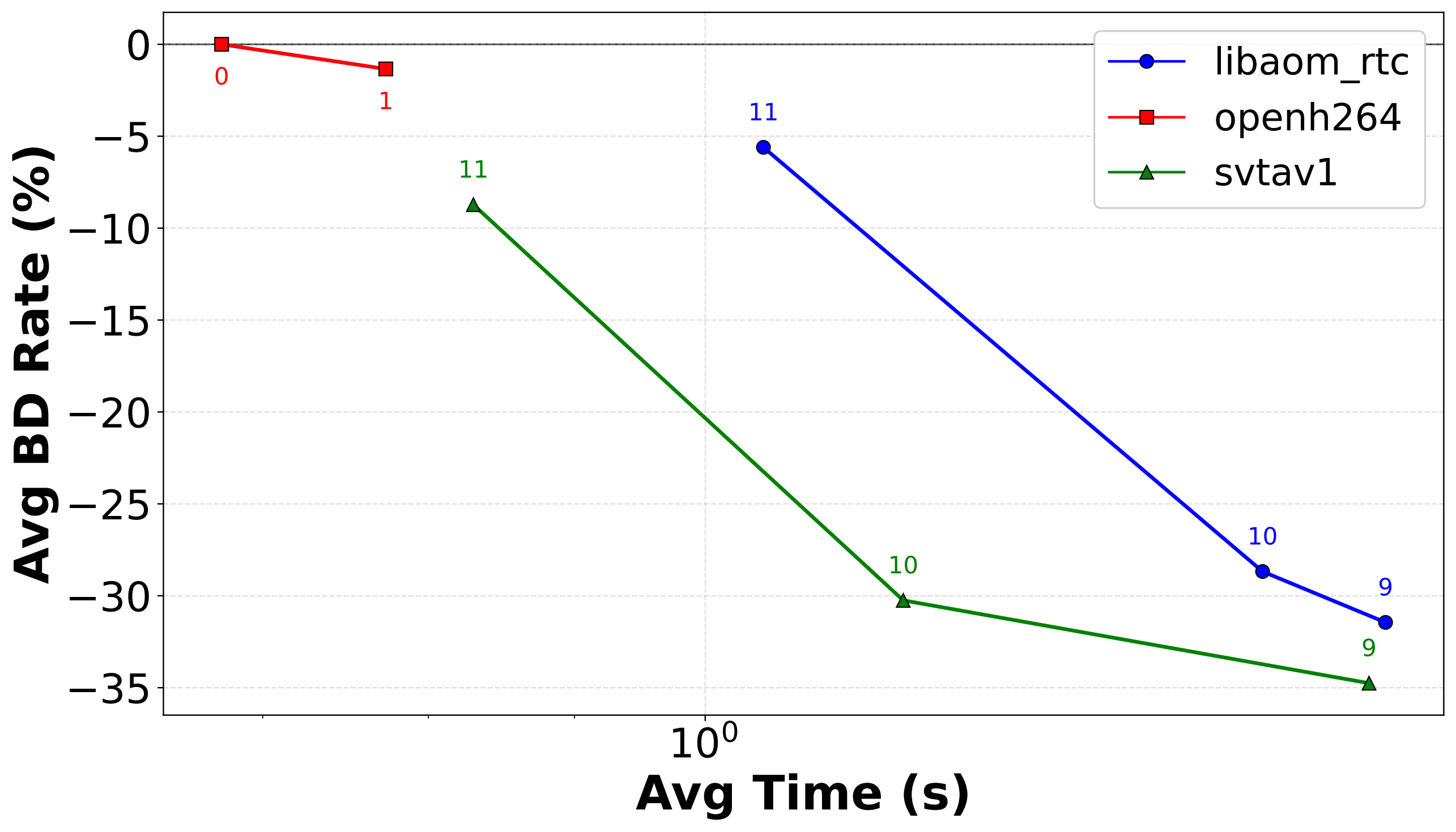}}
\caption{Encoding results (normalized to OpenH264 preset 0). Each dot represents an encoder preset.}
\label{fig1}
\end{figure}

Fig.~\ref{fig1} shows BD-rate savings versus encoding time for each encoder/preset combination, illustrating the Pareto frontier between efficiency and speed. The horizontal axis represents encoding time, while the vertical axis shows BD-rate savings. As expected, higher-complexity presets generally provide greater bitrate savings. The results show that SVT-AV1 preset 9 can achieve up to 35\% bitrate savings compared to the OpenH264 encoder.

\subsection{Analysis by Content Characteristics}
\begin{figure}[htbp]
\centerline{\includegraphics[width=\columnwidth]{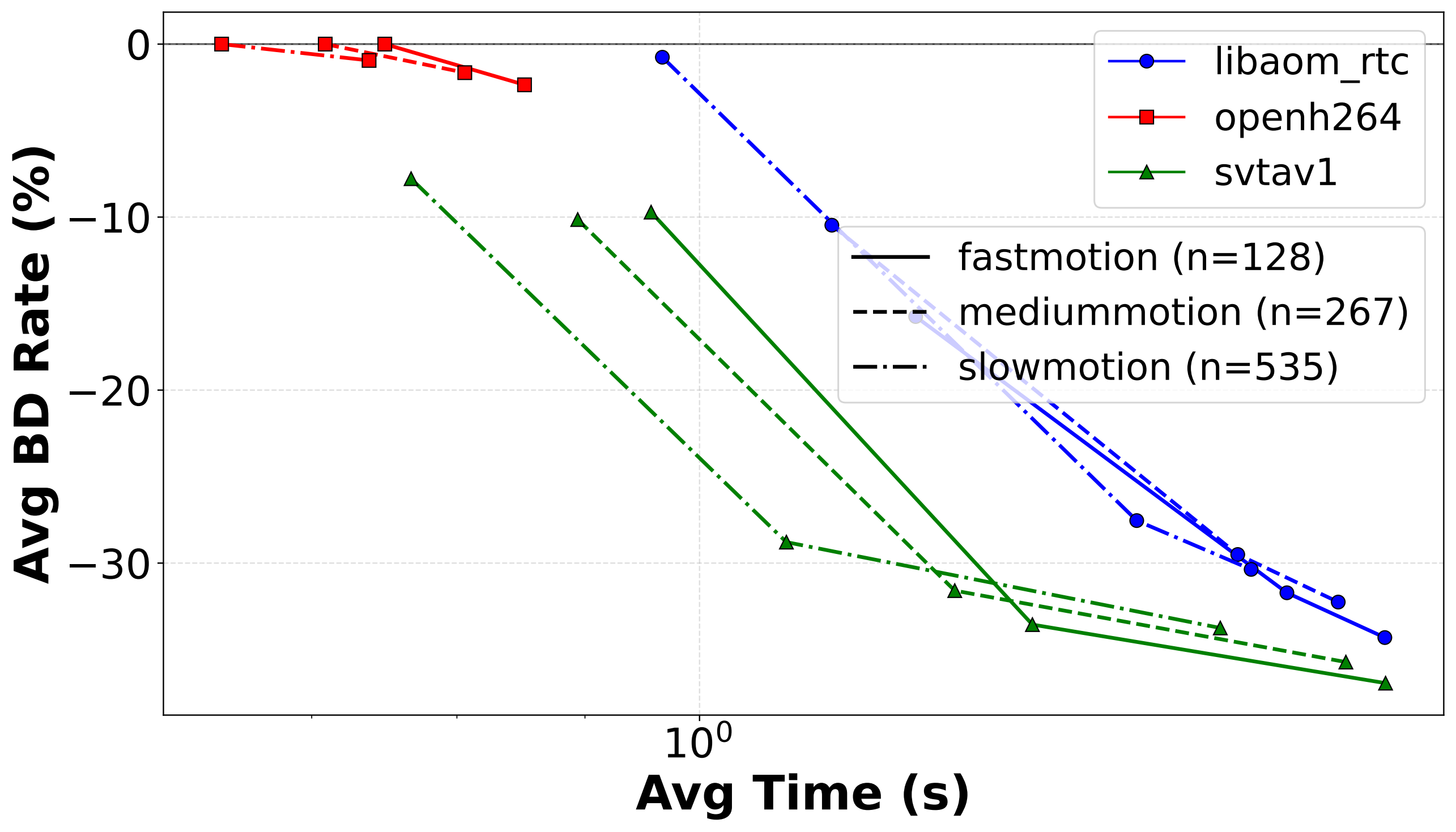}}
\caption{Encoding results for videos with different motion characteristics (normalized to OpenH264 preset 0), where n represents the number of video sequences. Each dot represents an encoding preset, consistent with Fig. 1.}
\label{fig2}
\end{figure}

We analyze encoder performance stratified by content characteristics to understand how each codec responds to the diverse conditions in VC-1K. Key findings:

\textbf{Motion impact}: Fig.~\ref{fig2} shows the results for videos with different motion characteristics. At the same encoder complexity, SVT-AV1 provides better bitrate savings on slow-motion videos than on fast-motion videos.

\begin{figure}[htbp]
\centerline{\includegraphics[width=\columnwidth]{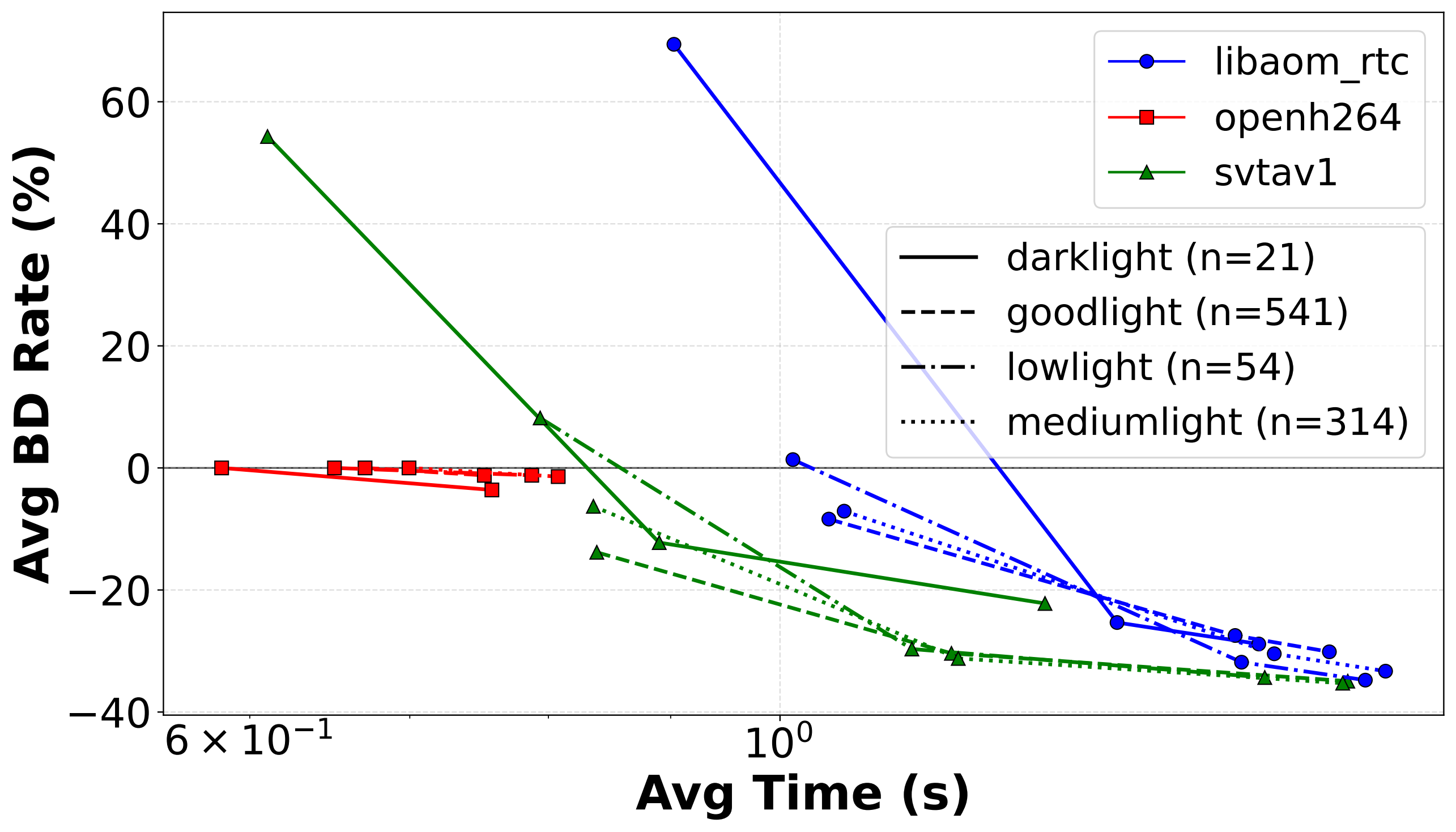}}
\caption{Encoding results for videos with different lighting characteristics (normalized to OpenH264 preset 0), where n represents the number of video sequences. Each dot represents an encoding preset, consistent with Fig. 1.}
\label{fig3}
\end{figure}

\textbf{Lighting impact}: Fig.~\ref{fig3} shows the results for videos with different lighting characteristics. The results indicate that dark/low-light sequences are more challenging for compression, largely due to sensor noise. For the AV1 encoders, the fastest presets perform noticeably worse on these sequences, suggesting that speed-oriented configurations may sacrifice rate–distortion–critical tools in a way that is suboptimal for noisy, low-SNR content. A plausible explanation is that both encoders are primarily optimized for cleaner, high-SNR content, and therefore generalize less effectively to low-light conditions.

\begin{figure}[htbp]
\centerline{\includegraphics[width=\columnwidth]{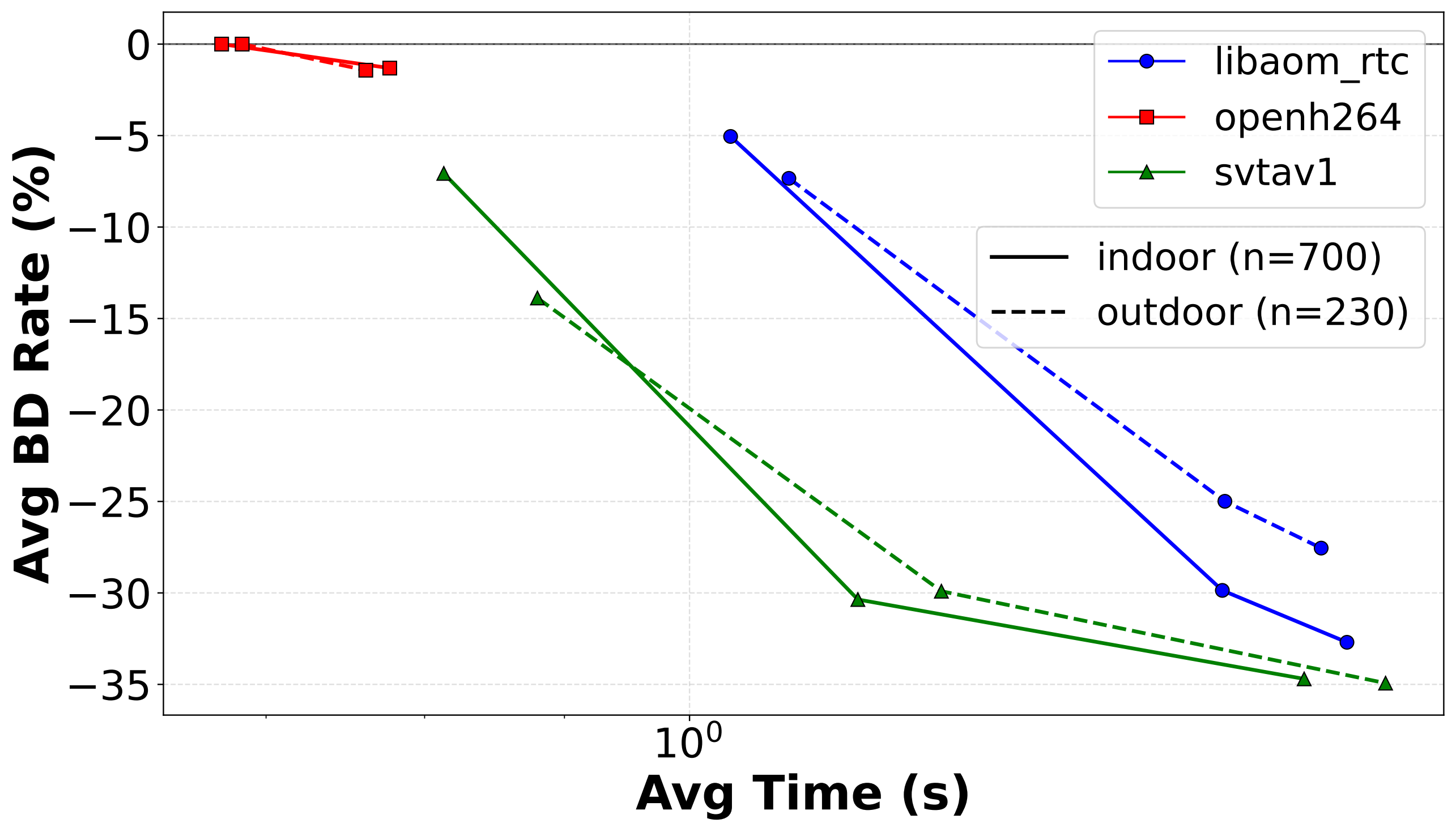}}
\caption{Encoding results for indoor and outdoor videos (normalized to OpenH264 preset 0), where n represents the number of video sequences. Each dot represents an encoding preset, consistent with Fig. 1.}
\label{fig4}
\end{figure}

\textbf{Indoor vs. Outdoor}: Fig.~\ref{fig4} shows the results for indoor and outdoor videos. The results indicate that outdoor videos are more difficult to compress and require more encoding time to achieve the same BD-rate savings.

\section{Super Resolution Evaluation}
To demonstrate the practical utility of our dataset for video enhancement applications, we conduct a comprehensive evaluation of super-resolution methods on VC-1K. We compare two representative approaches that exemplify different design philosophies: RealESRGAN \cite{b4}, a general-purpose super-resolution method, and GFPGAN \cite{b5}, which incorporates facial structural priors for face-specific enhancement.

Both methods are from the same open source project \cite{b4,b5} and share similar base architectures, allowing us to isolate the contribution of domain-specific optimization for facial content. For each video in our dataset, we extract the first 30 frames. We create low-resolution inputs by downsampling the original frames by a factor of 2 using bicubic interpolation, then apply both super-resolution methods to restore the original resolution. This controlled degradation allows for quantitative evaluation against ground-truth high-resolution frames. We employ three complementary metrics to assess super-resolution quality, such as PSNR (Peak Signal-to-Noise Ratio), SSIM (Structural Similarity Index), LPIPS (Learned Perceptual Image Patch Similarity). All metrics are computed on the RGB color space and averaged across all frames for each video.

Table~\ref{tab:sr} presents the quantitative comparison between RealESRGAN and GFPGAN on the proposed facial video dataset. Results are averaged across all videos in the dataset. Although the results show that GFPGAN has worse objective quality scores, its video outputs are actually clearer in the facial regions. We also observed an interesting phenomenon: when motion blur occurs in the video, GFPGAN can exhibit a deblurring effect. However, this may lead to a visual inconsistency where the background remains blurred while the face appears sharp. In addition, we found that when GFPGAN reconstructs faces from motion-blurred frames, it can sometimes alter textures, for example, changing the style of eyeglasses. Whether users prefer such effects is an open question worth further study. This type of issue cannot be uncovered using image datasets alone, and is one of the values provided by VC-1K.

\begin{table}[t]
\caption{Super-Resolution Comparison on VC-1K Facial Video}
\label{tab:sr}
\centering
%\footnotesize
\begin{tabular}{@{}lccc@{}}
\toprule
\textbf{Method} & \textbf{PSNR (dB)} & \textbf{SSIM} & \textbf{LPIPS} \\
\midrule
RealESRGAN (General)   & 33.81 & 0.91 & 0.13 \\
GFPGAN (Face-specific) & 31.31 & 0.89 & 0.16 \\
\bottomrule
\end{tabular}
\end{table}

\section{Conclusion}
We have presented VC-1K, a diverse and publicly available video test set of approximately 1,000 sequences designed specifically for evaluating technologies in real-time communication scenarios. The dataset provides controlled variation across device types, lighting conditions, motion levels, environments, and participant demographics, making it uniquely suited for comprehensive RTC benchmarking.

The encoder evaluation demonstrates SVT-AV1 preset 9 can achieve up to 35\% bitrate savings compared to the OpenH264 encoder. Dark/low-light video are more challenging for compression, largely due to sensor noise. The evaluation methodology, based on BD-Rate and CPU time measurements with open-source scripts \cite{b14}, is directly applicable to production video calling deployments and enables reproducible comparison of codec implementations.

The dataset, along with metadata and evaluation scripts, is publicly released to facilitate reproducible research in video communication codec development.

\section{Dataset Availability}
The VC-1K dataset, including all video sequences, metadata annotations, and evaluation scripts, is publicly available at: https://github.com/facebookresearch/VC-1K/. The dataset is released under CC-by-NC license (CC  Attribution-NonCommercial 4.0).

\end{document}